\documentclass[sigconf]{acmart}

\usepackage{siunitx}
\usepackage{hyperref} 
\usepackage{multirow}
\usepackage[linesnumbered,ruled,vlined,noend]{algorithm2e}
\usepackage{makecell}
\usepackage{pifont}
\usepackage{caption}
\usepackage{listings}
\usepackage{enumitem}
\definecolor{codered}{rgb}{0.8,0.2,0} % 定义红色
\definecolor{codegreen}{rgb}{0,0.6,0} % 定义绿色
\definecolor{codegray}{rgb}{0.2,0.2,0.2} % 定义灰色
\definecolor{codeblue}{rgb}{0,0.5,0.6} % 定义蓝色

\AtBeginDocument{%
  }

\copyrightyear{2025}
\acmYear{2025}

\setcopyright{cc}
\setcctype[4.0]{by}
\acmConference[Internetware 2025]{16th Symposium on Internetware}{June 20--22, 2025}{Trondheim, Norway}
\acmBooktitle{16th Symposium on Internetware (Internetware 2025), June 20--22, 2025, Trondheim, Norway}
\acmDOI{10.1145/3755881.3755918}
\acmISBN{979-8-4007-1926-4/25/06}

\begin{document}

% \title{\textsc{Tech-ASan}: \underline{T}wo-stag\underline{e} \underline{ch}eck for Address Sanitizer}
\title{\textsc{Tech-ASan}: Two-stage check for Address Sanitizer}
% \title{\textsc{Tech-ASan}:  Accelerate \underline{A}ddress \underline{San}itizer using \underline{T}wo-stag\underline{e} \underline{Ch}eck}
% \title{\textsc{Tech-ASan}: Accelerate Address Sanitizer using Two-Stage Check}
% \title{\textsc{Tech-ASan}: \underline{T}wo-stag\underline{e} \underline{Ch}eck \underline{A}ddress \underline{S}anitizer}
% \title{\textsc{DoubleASan}: \underline{D}ouble \underline{Ch}eck \underline{A}ddress \underline{S}anitizer}

\author{Yixuan Cao}
\orcid{0009-0006-6241-4251}
\affiliation{%
  \institution{College of Computer Science and\\ Software Engineering \\Shenzhen University}
  \city{Shenzhen}
  \state{Guangdong}
  \country{China}
}
\email{caoyixuan2019@email.szu.edu.cn}

\author{Yuhong Feng}
\authornote{Corresponding author.}
\orcid{0000-0002-7691-5587}
\affiliation{%
  \institution{College of Computer Science and\\ Software Engineering \\Shenzhen University}
  \city{Shenzhen}
  \state{Guangdong}
  \country{China}
  \postcode{518060}
}
\email{yuhongf@szu.edu.cn}

\author{Huafeng Li}
\orcid{0009-0006-9490-0449}
\affiliation{%
  \institution{College of Computer Science and\\ Software Engineering \\Shenzhen University}
  \city{Shenzhen}
  \state{Guangdong}
  \country{China}
}
\email{lihuafeng2020@email.szu.edu.cn}

\author{Chongyi Huang}
\orcid{0009-0004-7370-2377}
\affiliation{%
  \institution{College of Computer Science and\\ Software Engineering \\Shenzhen University}
  \city{Shenzhen}
  \state{Guangdong}
  \country{China}
}
\email{huangchongyi2020@email.szu.edu.cn}

\author{Fangcao Jian}
\orcid{0009-0001-4312-157X}
\affiliation{%
  \institution{College of Computer Science and\\ Software Engineering \\Shenzhen University}
  \city{Shenzhen}
  \state{Guangdong}
  \country{China}
}
\email{jianfangcao2023@email.szu.edu.cn}

\author{Haoran Li}
\orcid{0009-0007-6789-5573}
\affiliation{%
  \institution{College of Computer Science and\\ Software Engineering \\Shenzhen University}
  \city{Shenzhen}
  \state{Guangdong}
  \country{China}
}
\email{lihaoran2018@email.szu.edu.cn}

\author{Xu Wang}
\orcid{0000-0002-2948-6468}
\affiliation{%
  \institution{College of Computer Science and\\ Software Engineering \\Shenzhen University}
  \city{Shenzhen}
  \state{Guangdong}
  \country{China}
}
\email{wangxu@szu.edu.cn}

\begin{abstract}
Address Sanitizer (ASan) is a sharp weapon for detecting memory safety violations, including temporal and spatial errors hidden in C/C++ programs during execution. However, ASan incurs significant runtime overhead, which limits its efficiency in testing large software. The overhead mainly comes from sanitizer checks due to the frequent and expensive shadow memory access. Over the past decade, many methods have been developed to speed up ASan by eliminating and accelerating sanitizer checks, however, they either fail to adequately eliminate redundant checks or compromise detection capabilities. To address this issue, this paper presents \textsc{\underline{Tech}-ASan}, a \underline{t}wo-stag\underline{e} \underline{ch}eck based technique to accelerate ASan with safety assurance. First, we propose a novel two-stage check algorithm for ASan, which leverages magic value comparison to reduce most of the costly shadow memory accesses. Second, we design an efficient optimizer to eliminate redundant checks, which integrates a novel algorithm for removing checks in loops. Third, we implement \textsc{Tech-ASan} as a memory safety tool based on the LLVM compiler infrastructure. Our evaluation using the SPEC CPU2006 benchmark shows that \textsc{Tech-ASan} outperforms the state-of-the-art methods with 33.70\% and 17.89\% less runtime overhead than ASan and ASan\mbox{-}\mbox{-}, respectively. Moreover, \textsc{Tech-ASan} detects 56 fewer false negative cases than ASan and ASan\mbox{-}\mbox{-} when testing on the Juliet Test Suite under the same redzone setting.
\end{abstract}




\maketitle

\section{Introduction}
Programs written in memory-unsafe languages such as C and C++ often contain memory safety violations, which can be categorized into \textit{temporal errors} and \textit{spatial errors}. 
A temporal error occurs when code attempts to access a memory object after it has been freed, whereas a spatial error happens when the code reads or writes beyond the bounds of a valid memory object.

Memory safety violations are the root cause of many of today’s most severe vulnerabilities \cite{MSET2025}, which may lead to severe issues such as system crash, data breach, and hijacked execution when exploited \cite{Sok2013}. As reported in the 2023 CWE Top 10 KEV Weaknesses, use-after-free, heap-based buffer overflow, and out-of-bounds write rank 1st, 2nd, and 3rd among all weaknesses, respectively.\footnote{\url{https://cwe.mitre.org/top25/archive/2023/2023_kev_list.html}} 
In recent years, sanitizing for memory safety has become a widely adopted method for identifying vulnerabilities in software testing, especially within the realm of fuzzing \cite{10.1145/3658644.3690278,FloatZone2023}.
Motivated by numerous security incidents caused by memory safety violations in system software,
various sanitizers have been created to detect memory safety violations at runtime to help developers fix them, where Address Sanitizer (ASan) is the most popular tool due to its outstanding
\textit{capability} (detection of a wide spectrum of spatial errors and temporal errors), \textit{scalability} (ability to support industry-grade programs like operating system kernels and web browsers), and \textit{usability} (nearly zero configuration and seamless integration into mainstream compilers such as clang and gcc) \cite{ASAN2012,ASAN--2022}.

Technically, ASan allocates additional shadow memory, poisons/unpoisons the shadow memory at runtime to record the \textit{addressable} state of the memory, and checks the shadow memory before memory access to determine whether it is a memory safety violation. However, ASan brings approximately 1$\times$ runtime overhead to the tested program, which limits the efficiency of ASan in testing large software. Over the past decade, reducing the runtime overhead of ASan has attracted the interest of many researchers. Existing studies have shown that more than 80\% of the ASan's runtime overhead is introduced by sanitizer checks \cite{ASAN--2022}. To address this issue, the existing optimization solutions mainly focus on two aspects: \textit{check elimination} \cite{GWP-ASan2024,SANRAZOR2021,ASAP2015,BUNSHIN2017,PartiSan2018,ASAN--2022,doubletake2016} and \textit{check acceleration} \cite{FloatZone2023,giantsan2024}.

\textbf{Check elimination. }As the official ASan documentation\footnote{\url{https://github.com/google/sanitizers/wiki/AddressSanitizerCompileTimeOptimizations}} states, finding all bugs does not require to instrument all memory accesses, sanitizer checks for memory with safety assurance can be eliminated. Existing check-eliminating methods can be categorized into \textit{performance-driven} and \textit{security-driven} methods. Performance-driven methods \cite{doubletake2016,GWP-ASan2024,ASAP2015,PartiSan2018} eliminate sanitizer checks to meet performance constraints but do not guarantee safety.
DoubleTake \cite{doubletake2016} uses canaries to overwrite unaddressable memory and divides the program runtime into several epochs, checking whether the canary has been tampered with at the end of each epoch. However, this design can only detect write vulnerabilities, but not read vulnerabilities. In addition, DoubleTake only partially overwrites the freed heap memory with canaries, so it will miss use-after-free in unprotected areas. GWP-ASan \cite{GWP-ASan2024} randomly protects heap objects with guard pages and ignores most of the others. ASAP \cite{ASAP2015} removes sanitizer checks on “hot” code that is more frequently executed. PartiSan \cite{PartiSan2018} follows a performance-driven metric to remove sanitizer checks.

Security-driven methods \cite{SANRAZOR2021,ASAN--2022} are designed to eliminate redundant sanitizer checks to reduce runtime overhead with safety assurance. 
SANRAZOR \cite{SANRAZOR2021} combines static patterns and dynamic patterns to identify and remove redundant sanitizer checks, but it still cannot ensure the removed checks are indeed redundant due to its unsound patterns. Also, SANRAZOR needs user input for profiling. Therefore, SANRAZOR degrades ASan's capability and usability. Finally, ASan\mbox{-}\mbox{-} \cite{ASAN--2022} uses static analysis to eliminate redundant checks while ensuring safety at compile time, which is still the state-of-the-art (SOTA) method to accelerate ASan. However, ASan\mbox{-}\mbox{-} still has room for further optimization: First, it fails to speed up the sanitizer check. Second, it fails to fully remove checks within loops. Although checks in loops account for about 45\% of the overhead
introduced by all ASan checks \cite{ASAN--2022}, eliminating redundant checks in loops remains unsolved.
In summary, existing check elimination methods either eliminate checks without safety assurance or fail to adequately eliminate checks on security access.

\textbf{Check acceleration. }In recent years, some works have attempted to accelerate checks to reduce the overall runtime overhead of the sanitizer, but they compromise the capabilities of ASan. FloatZone \cite{FloatZone2023} replaces comparison-based checks with floating-point underflow exception-based checks to enable higher instruction-level parallelism, and cancels shadow memory to reduce cache miss rate. But FloatZone introduces false positives because metadata is no longer isolated in shadow memory. GiantSan \cite{giantsan2024} introduces history
caching and region checking to achieve fast operation-level
protection. However, our experimental evaluation shows that it still misses memory safety violations in the Juliet Test Suite \cite{black2018juliet} and in the real world that can be detected by ASan. In summary, both FloatZone and GiantSan reduce ASan's detection capability.

From the above analysis, we state the problem as: How to further reduce the runtime overhead while maintaining the capability, scalability, and usability of ASan? To address it, we propose \textsc{Tech-ASan}, a novel ASan optimization method.
First, \textsc{Tech-ASan} introduces a two-stage checker: when checking a memory access instruction, the fast check stage determines whether the accessing location contains a magic value. If so, the slow check stage checks the metadata of shadow memory to determine whether the memory is addressable. Since the probability of triggering the slow checker is very low and the slow check stage accesses shadow memory infrequently, the check is accelerated.
Second, \textsc{Tech-ASan} proposes an effective optimizer to identify and eliminate redundant checks, which integrates an original algorithm for removing checks in loops.

The contribution of this paper can be summarized as follows:
\begin{itemize}
    \item We design a novel and fast two-stage sanitizer check algorithm for ASan, which does not rely on loading shadow bytes in most cases.
    \item We propose an optimizer to eliminate redundant sanitizer checks with safety assurance, which integrates a novel algorithm for removing checks within the loop.
    \item We implement \textsc{Tech-ASan} as a memory safety tool based on LLVM compiler infrastructure \cite{LLVM2004}.
    \item Our comprehensive evaluation experiments show that \textsc{Tech-ASan} outperforms the SOTA methods with 33.70\% and 17.89\% less runtime overhead than ASan and ASan\mbox{-}\mbox{-}, respectively, while maintaining the advantages of ASan.
\end{itemize}
The rest of the paper is organized as follows. Section \ref{sec:Background} reviews the technical background of ASan. Section \ref{sec:Approach} presents our original optimization approach for ASan. Section \ref{sec:Evaluation} describes the performance evaluation. Section \ref{sec:related works} reviews the related work, and finally Section \ref{sec:conclusion}
concludes the paper.

\section{Background} \label{sec:Background}
This section briefly reviews the technical background of ASan from three aspects: \textit{shadow memory}, \textit{redzone}, and \textit{runtime check}.

\subsection{Shadow Memory}
\begin{figure}[tb]
  \centering
  \includegraphics[width=\linewidth, page=1]{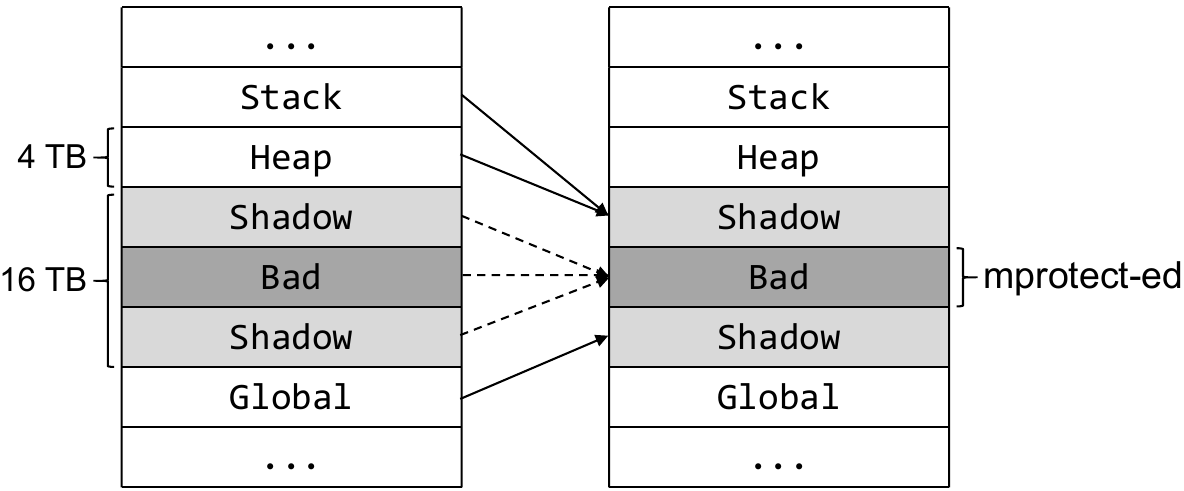}
  \caption{Shadow memory model in a 64-bit address space}
  \label{fig:shadow}
\end{figure}

As shown in Figure \ref{fig:shadow}, ASan utilizes a shadow memory model to support sanitizer checks.
By default, ASan spares one-eighth of the virtual address space as shadow memory, where each shadow byte records the status of eight bytes used by the application.
Given the application memory address \texttt{Addr}, which can be located in stack, heap, or global regions, the corresponding address of the shadow byte is computed as \texttt{(Addr \mbox{>}\mbox{>} 3) + Offset}, where the \texttt{Offset} is a constant that must be chosen statically at the compiling time for every platform. With the same address computation way, addresses in the shadow memory can be mapped into the $Bad$ region, which is protected via page protection.

In shadow memory, each byte is encoded with the following regular: 0 means that the entire 8-byte corresponding application memory region is addressable; \texttt{k} $ (1\le \texttt{k} \le 8)$ means that the first \texttt{k} bytes are addressable and the last (8 - \texttt{k}) bytes are not; any negative value means that the entire 8-byte word is unaddressable. In the last scenario, different negative values are used to indicate different types of memory safety violations, which help developers to locate and fix bugs.

\subsection{Redzone}
\begin{figure}[tb]
  \centering
  \includegraphics[width=\linewidth]{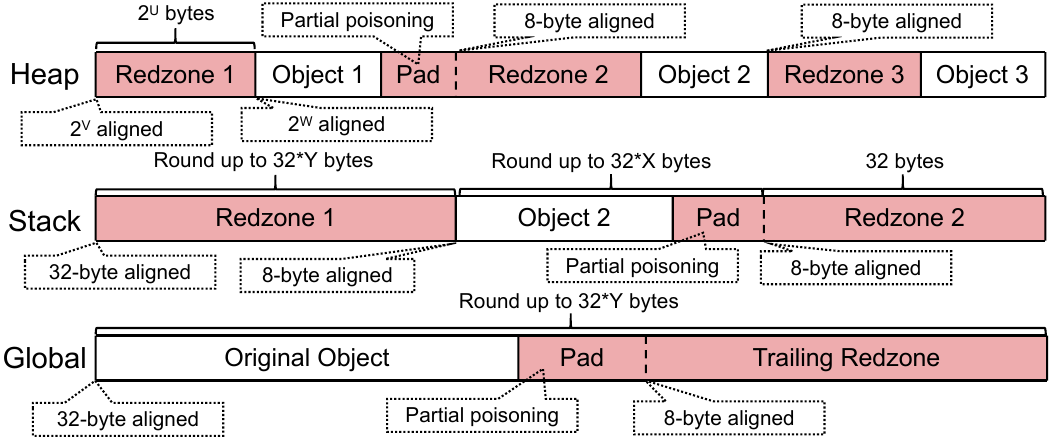}
  \caption{ASan's redzones in heap, stack, and global regions, where \texttt{U}, \texttt{V}, \texttt{W}, \texttt{X}, and \texttt{Y} represent positive integers.}
  \label{fig:redzone}
\end{figure}

Figure \ref{fig:redzone} shows that ASan places redzones before and after each memory object in interesting memory regions, including stack, heap, and global \cite{ASAN2012,ASAN--2022}. When allocating a memory object, ASan sets the object itself as addressable and poisons its redzones as unaddressable to detect spatial errors, e.g., heap-buffer-overflow and stack-buffer-overflow. When freeing a memory object, ASan poisons the freed memory object as unaddressable to detect temporal errors, e.g., heap-use-after-free and double-free.
Note that poisoning operators are to change the shadow byte but not the application memory.

\textbf{Heap. }There are several heap allocation functions and operators provided by the standard C and C++ languages, i.e., \texttt{malloc()}, \texttt{calloc()}, \texttt{mmap()}, \texttt{new()}, and \texttt{new[]()}. ASan replaces them with customized functions. When a heap object is allocated, ASan places a redzone both before and after the buffer. The size of the redzone is a power of two, which can range from 16 to 2,048 according to the object size.
When an object is freed by the functions or the operators provided by the standard C and C++ languages, i.e., \texttt{free()}, \texttt{munmap()}, \texttt{delete}, and \texttt{delete[]}, the freed object is poisoned and quarantined into a 256MB queue to detect heap-use-after-free and double-free.

\textbf{Stack. }The Stack objects can be allocated statically or allocated dynamically by the C standard function \texttt{alloca()}. For each stack object, ASan allocates and poisons the left redzone (32 bytes) and the right redzone (32 bytes plus up to 31 bytes for alignment) when entering a function.

\textbf{Global. }The global region manages global variables, static variables, and constants in programs written in the C and C++ languages, which corresponds to \texttt{.data} and \texttt{.bss} segments.
ASan places a trailing redzone at the right of each global object. The redzone size is the larger of 32 bytes or one-fourth of the object size, plus a padding size if the object size is not 8-byte aligned.
Finally, the size of the object plus the redzone is rounded up to a multiple of 32 bytes.
All the redzones for global objects are poisoned when initializing the process. Since global objects are never to be freed, no temporal error occurs in the global region, which provides an optimization opportunity for eliminating unnecessary sanitizer checks.

\subsection{Runtime Check}
ASan instruments every memory access instructions, including load and store instructions in Intermediate Representation (IR) code, to check whether an access is addressable. There are two ways to implement this: inserting an inline instruction sequence or a function call. In order to compromise between runtime performance and code size, ASan inserts inline instructions by default to reduce runtime overhead, but when the number of instrumentation reaches a threshold, the check function is called. 

Depending on the size of the memory access, the check works differently. When instrumenting an 8-byte memory access, ASan computes the address of the corresponding shadow byte, loads that byte, and checks whether it is 0:
\begin{lstlisting}%[caption={ASan's check on 8-byte access}, label={asan8byteCheck}]

ShadowAddr = (Addr >> 3) + Offset;
if (*ShadowAddr != 0)
    ReportAndCrash(Addr);
\end{lstlisting} 
where \texttt{Offset} is a large platform-dependent constant. When instrumenting an N-byte memory access, where N = 1, 2, or 4, ASan checks if the first \texttt{k} bytes in the 8-byte word are addressable:
\begin{lstlisting}%[caption={ASan's check on N-byte access (N = 1, 2, 4, or 8)}, label={asanNbyteCheck}]

ShadowAddr = (Addr >> 3) + Offset;
k = *ShadowAddr;
if (k != 0 && ((Addr & 7) + AccessSize > k))
    ReportAndCrash(Addr);
\end{lstlisting}

ASan also intercepts standard C/C++ library functions that may cause memory errors (e.g., \texttt{memset()} and \texttt{strcpy()}). When such a function is called, ASan checks the shadow memory using an optimized function \texttt{\_\_asan\_region\_is\_poisoned()} to determine whether the accessed memory region is entirely addressable.

ASan's sanitizer checks rely on obtaining the corresponding shadow address through shift and addition calculations, loading the shadow byte, and determining whether it is a memory safety violation via comparison that may be more than once.
Frequent access to shadow memory destroys the locality of the program, which increases runtime overhead.
Our design focuses on how to complete the sanitizer check with only a fast comparison in most cases, without introducing expensive load shadow byte operations.

\section{Approach}\label{sec:Approach}

\begin{figure}[tb]
  \centering
  \includegraphics[width=\linewidth]{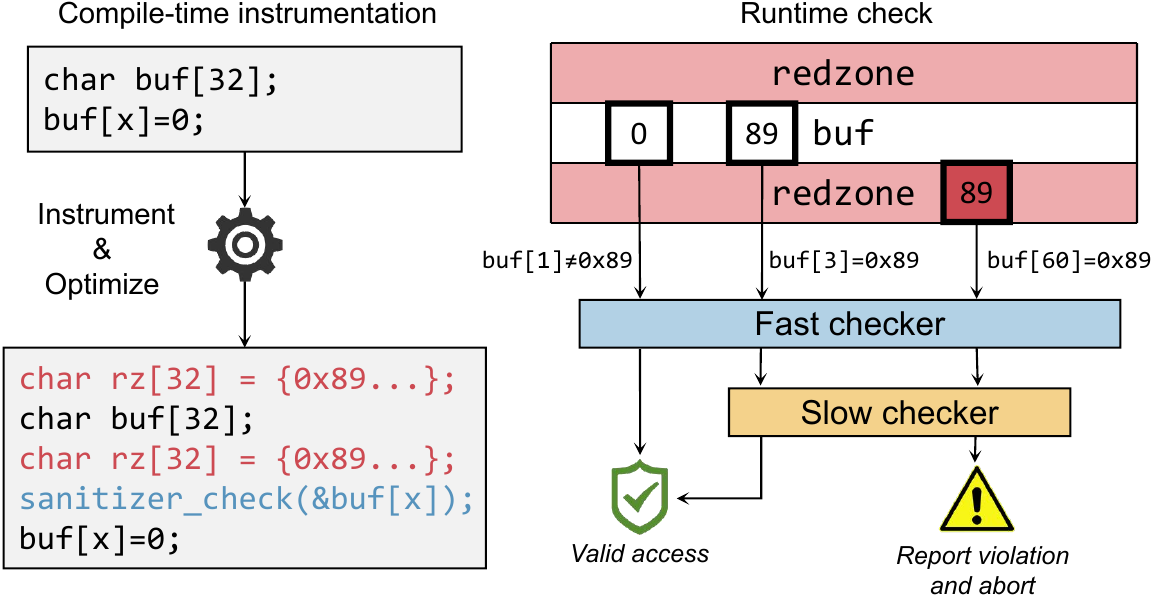}
  \caption{ The framework of \textsc{Tech-ASan}}
  \label{fig:overview}
\end{figure}

This section elaborates on how \textsc{Tech-ASan} is designed to accelerate ASan. Figure \ref{fig:overview} overviews \textsc{Tech-ASan}, which consists of a compile-time instrumentation phase and a runtime check phase.
In the compile-time instrumentation phase, \textsc{Tech-ASan} takes the program source code as its input, and outputs an instrumented executable program. Specifically, the program source code is first compiled into LLVM IR \cite{LLVM2004}, and then instrumented to manage redzones and inject a magic value, as well as check violations (cf. Section \ref{sec:Instrumentation}). Unlike ASan, we not only poison and unpoison the shadow memory, but also poison and unpoison the redzones and the freed memory regions using the magic value, i.e., \texttt{0x89}. Since not all checks are necessary, \textsc{Tech-ASan} utilizes an optimizer to reduce runtime overhead by eliminating redundant checks (cf. Section \ref{sec:Optimization}).

In the runtime check phase, \textsc{Tech-ASan} executes normally if no memory safety violation is detected (e.g., \texttt{buf[1]} in Figure \ref{fig:overview}), otherwise it reports a violation and aborts the process immediately. Specifically, we design a two-stage checker where every sanitizer check is divided into two stages (cf. Section \ref{sec:Two-stage Check}). In the fast check stage, \textsc{Tech-ASan} checks whether there is a magic value in the access location, serving as a fast filtering mechanism. When a magic value is detected during the fast check stage, the slow check stage is started for a more precise verification: If the slow checker does not detect an invalid memory access, the program continues execution (e.g., \texttt{buf[3]} in Figure \ref{fig:overview}), otherwise a violation is reported and the program is terminated (\texttt{buf[60]} in Figure \ref{fig:overview}).

\subsection{Instrumentation}
\label{sec:Instrumentation}
This section provides details on how \textsc{Tech-ASan} instruments the program under test.
\textsc{Tech-ASan} first compiles the program under test from source code into LLVM IR and then proceeds with instrumentation. Similar to post compiler-based methods  \cite{FloatZone2023,giantsan2024,UAFSan2021,ASAN2012,ASAN--2022}, \textsc{Tech-ASan}'s instrumentation operations are performed at the IR level, therefore, the discussions in this paper do not involve memory objects managed by third-party libraries with inaccessible source code.
\textsc{Tech-ASan} reuses the ASan's instrumentation infrastructure as reviewed in Section \ref{sec:Background}, and we focus on the two main differences between \textsc{Tech-ASan} and ASan, namely \textit{magic value injection} and \textit{memory safety check}.

\textbf{Magic value injection.}
As illustrated in Figure \ref{fig:redzone}, ASan places redzones on both sides of heap, stack, and global memory objects to detect spatial errors. When a memory object is allocated, ASan poisons shadow bytes corresponding to the redzones, and 
unpoisons shadow bytes corresponding to the memory object regions to detect spatial errors.
When a memory object is freed, ASan poisons the shadow bytes corresponding to the memory region of that object to detect temporal errors. \textsc{Tech-ASan} adopts the shadow memory management approach of ASan, and goes a step further by filling the redzones and the freed memory regions with a magic value, as shown in Figure \ref{fig:Two-stage-checks}.
Considering the allocated memory objects in ASan 
(including reused ones)
inherently come with initial values, unlike the management of shadow memory, the magic value injection process in \textsc{Tech-ASan} only requires poisoning and does not involve explicit unpoisoning. As a result, the additional runtime overhead introduced by magic value injection is minimal.

\textbf{Memory safety check.}
Similar to \cite{ASAN2012, FloatZone2023,giantsan2024,ASAN--2022}, \textsc{Tech-ASan} instruments memory access instructions and functions at compile time to check for memory safety violations. First, for all functions in the program, \textsc{Tech-ASan} sequentially traverses all instructions to identify \textit{the interesting memory access instructions}, such as {Load} instructions and {Store} instructions.
For fair comparison, we ensure the coverage of the interesting memory access instructions exactly matches both ASan and ASan\mbox{-}\mbox{-}.
Then, for each identified memory access instruction, we insert a two-stage checking logic (cf. Section \ref{sec:Two-stage Check}). Additionally, we design an optimizer to eliminate redundant checks (cf. Section \ref{sec:Optimization}).

Second, \textsc{Tech-ASan} replaces standard library functions in C/C++ that may trigger memory safety violations, such as \texttt{memcpy()} and \texttt{strcat()}, with customized versions. Notably, existing methods \cite{ASAN2012,giantsan2024,ASAN--2022}, including ASan in the latest version of LLVM (LLVM 20.1.0)\footnote{\url{https://github.com/llvm/llvm-project/releases/tag/llvmorg-20.1.0}}, overlook functions that operate on \texttt{wchar\_t} strings, such as \texttt{wcscpy()}. 
Considering that memory errors can also occur in these functions \cite{black2018juliet}, \textsc{Tech-ASan} supports detecting them to improve detection capability. In addition, \textsc{Tech-ASan} replaces functions and operators such as \texttt{free()}, \texttt{munmap()}, \texttt{delete}, and \texttt{delete[]} with the corresponding customized functions to detect double-free. The checking logic within the aforementioned customized  functions remains consistent with that of ASan and ASan\mbox{-}\mbox{-}.

\subsection{Two-stage Check}
\label{sec:Two-stage Check}
\begin{figure}[tb]
  \centering
  \includegraphics[width=\linewidth]{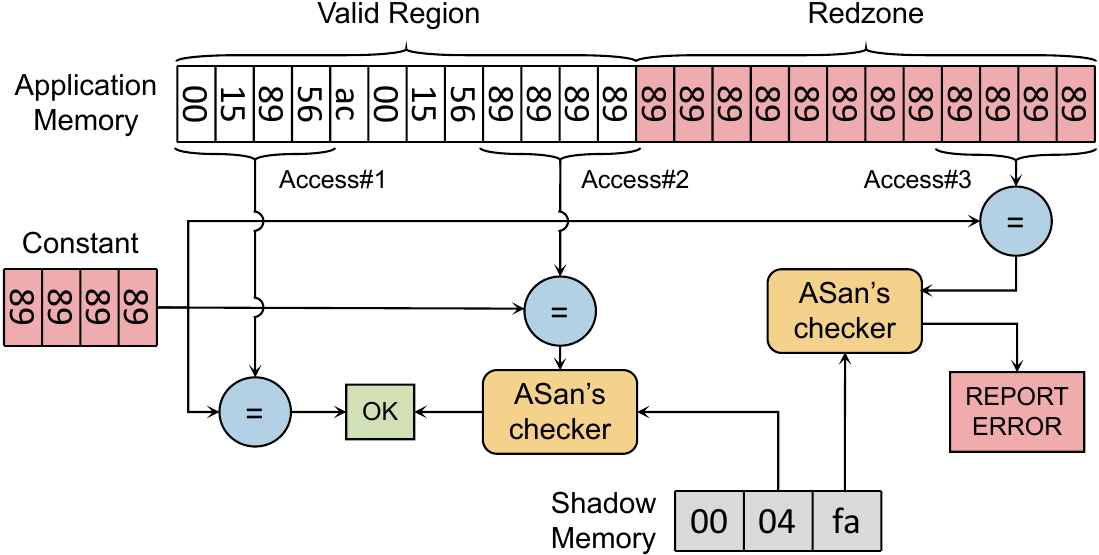}
  \caption{Examples of two-stage checks}
  \label{fig:Two-stage-checks}
\end{figure}
\textsc{Tech-ASan} utilizes a two-stage checker to implement the fast check for all interesting memory accesses, which consists of a fast check stage and a slow check stage. In the fast check stage, \textsc{Tech-ASan} checks whether the value of the accessed memory is equal to a predefined \texttt{N}-byte constant magic value \texttt{MAGIC\_VALUE\_N}, where \texttt{N} is 1, 2, 4, or 8. If not, the access is considered valid, and program execution continues as normal. If they are equal, the access is potentially located in a freed memory region or a redzone. 
However, since the program itself may legitimately use the same magic value, the access cannot be immediately classified as an invalid access. In such cases, \textsc{Tech-ASan} proceeds to the slow check stage for further verification to eliminate false positives. In the slow check stage, \textsc{Tech-ASan} adopts the same check logic as ASan, verifying the validity of the access by checking the corresponding shadow byte. Figure \ref{fig:Two-stage-checks} presents three cases when using the two-stage checking of \textsc{Tech-ASan}:
\begin{itemize}
    \item Access\#1 (valid): The value at the accessed location is not a magic value, so the fast check stage determines it as a valid access.
    \item Access\#2 (valid): The value at the accessed location is a magic value, causing the fast check stage to flag it as potentially invalid and triggering the slow check stage. The slow stage, which uses ASan's native checker, reads the shadow byte (\texttt{0x04}) and confirms it as a valid access.
    \item Access\#3 (invalid): The fast stage first detects a magic number at the accessed memory location, and the slow check stage further identifies a negative shadow byte, confirming it as an invalid memory access.
\end{itemize}
This two-stage checker ensures efficient detection capability of memory violations while efficiently reducing runtime overhead.

To further reduce the runtime overhead of \textsc{Tech-ASan}, we adopt different strategies for checking store and load instructions. For store instructions accessing 1-, 2-, 4-, or 8-byte memory regions, \textsc{Tech-ASan} inserts the following checking logic \textbf{before} the instruction:
% \begin{mdframed}[style=codebox]
\begin{lstlisting}%[caption={Two-stage check for a store instuction}, label=lst:store_check]

if (*p == MAGIC_VALUE_N)          // Fast check
    asan_check(p);                // Slow check
*p = 1;                  // A store instruction
\end{lstlisting}
% \end{mdframed}
Only by a combination of load and compare instructions (Line 1) can \textsc{Tech-ASan} check a store instruction (Line 3) in most cases, i.e., when the slow check is not triggered. Since the load instruction (Line 1) and the store instruction (Line 3) access the same address, no additional cache miss is introduced. Our statistical analysis of SPEC CPU2006 benchmarks confirms that 85\% of \textsc{Tech-ASan}'s interesting memory access instructions target load instructions. Although the above check logic is efficient, the \texttt{*p} appears twice, resulting in one additional memory access compared to the original program.
For load instructions accessing 1-, 2-, 4-, or 8-byte memory regions, 
\textsc{Tech-ASan} inserts the checking logic \textbf{after} the instruction:
% \begin{mdframed}[style=codebox]
\begin{lstlisting}%[caption={Two-stage check for a load instuction}, label=lst:load_check]

foo = *p;                 // A load instruction
if (foo == MAGIC_VALUE_N)         // Fast check
    asan_check(p);                // Slow check
\end{lstlisting}
% \end{mdframed}
In Line 2, we reuse the result of \texttt{*p} to avoid introducing additional memory access overhead. Note that if we do not explicitly reuse the result of \texttt{*p} when inserting check logic, LLVM's backend does not automatically reuse it. Therefore, unlike recent methods \cite{ASAN--2022,FloatZone2023} that require inserting an extra load instruction to read metadata from memory before the origin load instruction, we directly reuse the value from the origin load instruction, saving a load instruction.

The two-stage checker design of \textsc{Tech-ASan} is based on two  fundamental observations. First, the majority of memory accesses are valid in real-world software testing. Second, although ASan's per-access shadow memory validation is accurate, it can reduce cache hit rates and degrade overall performance. Guided by the principle of locality in program execution, \textsc{Tech-ASan}'s design philosophy argues that it is unnecessary to perform slow checks for every memory access as ASan does. For most valid accesses, executing only the fast-stage check is sufficient to efficiently filter out valid accesses and significantly accelerate checks. For suspicious accesses, \textsc{Tech-ASan} relies on slow-stage shadow memory checks to eliminate false positives, thus achieving a balance between runtime performance and detection capability.

Due to the low probability of triggering the slow check, \textsc{Tech-ASan} implements the slow checker as a function call rather than an inline instruction version. This is because the overhead of a function call is acceptable in this context. Additionally, this approach helps reduce the size of the binary file, further improving the usability of \textsc{Tech-ASan}.

% \subsection{Check Acceleration}

\subsection{Optimization}
\label{sec:Optimization}
Since there is no need to instrument on all memory accesses, \textsc{Tech-ASan}'s optimizer is designed to eliminate redundant checks. 
Among the most advanced redundant check elimination techniques, the most basic and important idea is \textit{removing unsatisfiable checks} on stack and global memory objects. Let us take the code in Listing \ref{lst:example} as an example. ASan's native code has a basic implementation. For the memory access in Line 4, ASan determines that the offset of the accessed location is less than the length of the memory object, ensuring that a buffer overflow cannot occur. As a result, it removes the check in Line 3. Based on this philosophy, ASan\mbox{-}\mbox{-} has further optimizations: For the memory access in Line 7, ASan\mbox{-}\mbox{-} uses data flow analysis to determine that the access in Line 7 is only triggered when the condition in Line 5, i.e., \texttt{i < 20}, is satisfied. Since the condition in Line 5 already ensures that the access in Line 7 cannot cause a buffer overflow, the check in Line 6 is removed.

\begin{lstlisting}[caption={An example of redundancy elimination}, label=lst:example]
int buf[20];
unsigned int i = input();
sanitizer_check(buf+10);     // Removed by ASan
buf[10] = 0;
if (i < 20) {
    sanitizer_check(buf+i);// Removed by ASan--
    buf[i] = 0;
}
for (unsigned int j = 0; j < 20; j++) { 
    sanitizer_check(buf+j);    // Removed by us
    buf[j] = input();
    sanitizer_check(buf);      // Removed by us
    buf[0] = 1;
}
\end{lstlisting}
% \end{mdframed}

However, removing unsatisfiable checks of ASan\mbox{-}\mbox{-} becomes ineffective in loops (Lines 11 and 13) because its data flow analysis struggles to handle PHI nodes in LLVM IR adequately. The PHI node is an IR instruction that selects incoming values from the different predecessor basic blocks in the form of static single assignment (SSA) rule. However, PHI nodes will also be obstacles for data flow analysis.
% Due to the static single assignment (SSA) rule, the PHI node is introduced to resolve problems that occurs when multiple paths execute (if-else branches) assign  different values to the same variable in the program. 
% When programs execute branching operations (like if-else statements), different code paths may assign different values to the same variable. Phi nodes resolve this conflict at control flow merge points. 
% Phi nodes play an important role in LLVM because of static single assignment (SSA) rule of IR. This rule allows only one value assignment per variable. When programs execute branching operations (like if-else statements), different code paths may assign different values to the same variable. Phi nodes resolve this conflict at control flow merge points. LLVM can identify natural loop structures. Each natural loop has one unique entry point called Header. The loop tail connects back to Header through backward edges, creating repetitive execution conditions. Phi nodes manage loop indexes in LLVM. At each loop iteration start, they choose initial values. These values come from previous calculation results. In subsequent iterations, they choose indexes from back edges.
Although ASan\mbox{-}\mbox{-} introduces specific extra optimizations targeting loop scenarios to mitigate this issue, it inevitably adds new operations. As a result, in our tests, the benefits are not significant.
As highlighted in \cite{ASAN--2022}, checks in loops account for about 45\% of the overhead
introduced by all ASan checks. Therefore, safely eliminating redundant checks inside loops remains an important and unresolved problem.

\textsc{Tech-ASan} first integrates the unsatisfiable check removal techniques from both ASan and ASan\mbox{-}\mbox{-}. Then, for removing Lines 10 and 12 in Listing \ref{lst:example},
Algorithm \ref{alg:Removing Checks} is proposed to eliminate redundant checks inside loop bodies {via} static analysis, i.e., {checks for} $a[index]$, where $index$ is a variable or a constant and $a$ is the base address of a stack or global object. 
% It mainly targets single-level loops which can be standardized. Focusing on single-level loops avoids the compile-time overhead caused by analyzing multi-level loops.
% Moreover, optimizing single-level loops is sufficient for most real-world applications.  
% In LLVM's IR, one memory access can mainly undergo two steps. First, the \texttt{getElementPtr} (GEP) instruction calculates the exact address of a memory location by a base address and index offsets. Second, \texttt{load} or \texttt{store} instructions use this computed address to access memory. In LLVM, natural loops that have a clear structure with a single entry (\texttt{Header}) and explicit back edges pointing to \texttt{Header}, can be identified and analyzed by LLVM.\footnote{\url{https://llvm.org/docs/LoopTerminology.html}}. If a memory access occurs in the natural loop, it means LLVM can describe this memory access with three aspects,  
% The proposed method works by tracing backward each user $u_i$ of $index$ variable in memory accesses with use-def chains to analyze access validity.
Here we define the size of the memory object as $size$ and a constant as $c$. A memory access $m_i$ is safe (thus satisfying check elimination) if one of the following conditions holds (Lines 1-{15}):

\begin{enumerate}
    % \item The analysis finds $index$ is always within a safe range $[0, c]$ where $0 \le c < size$, where $size$ represents the size of the memory object. 
    % \item The memory access instruction $m_i$ dominates the compare instruction $u_q$ or $u_q$ dominates $m_i$.
    \item If the index is a constant, the analysis finds $index$ is always within a safe range $[0, c]$ where $0 \le c < size$.
    \item If the index is a variable, not only does $index$ need to satisfy the safe range as above, but also the memory access instruction $m_i$ dominates the compare {instruction} {$cmp$} or {$cmp$} dominates $m_i$.
    % (either $m$ dominates $u_i$ or vice versa) (Line 8). 
\end{enumerate}
Due to loop semantics, $index$ remains constrained by the initial index value and the compare instructions served as the loop termination condition. If the initial index value is in bounds, even if the access instruction $m_i$ dominates the compare instruction {$cmp$} in control flow, in the subsequent iteration, the access instruction $m_{i+1}$ is still bounded by the compare instruction {$cmp$} of the previous iteration (Line {13}). 

% For each memory access instruction $m_i$ in function $\mathcal{F}$, the algorithm finds matched getelementptr $gep$ instruction which contains base address and offset indexes. Every index $g_j$ of $gep$ needs to be verified its safety. If $g_j$ is comes from a PHI node $p$, it means $g_j$ is a induction variable, then all incoming values from different branches would be explored. In this context, safety of $g_j$ is guaranteed only when each incoming value $v_k$ falls within the safe range. If $g_j$ is not a constant nor a variable from a PHI node, it indicates that $g_j$ is loop invariant and needs to be comfirmed whether within the safe range alse well. 

Algorithm \ref{alg:Removing Checks} scans each memory access instruction $m_i$ in the function $\mathcal{F}$ to find its corresponding getelementptr instruction $gep$ (Lines {18-22}). Each getelementptr instruction $gep$ consists of a base address and offset indexes, where each index $g_j$ needs to be found whether it is in bounds {(Lines 24-39)}. For each $g_j$:
% LLVM IR follows the static single assignment (SSA) rule. This rule allows only one value assignment per variable. When programs execute branching operations (like if-else statements), different code paths may assign different values to the same variable. Phi nodes resolve this conflict at control flow merge points. LLVM can identify natural loop structures. Each natural loop has one unique entry point called Header. The loop tail connects back to Header through backward edges, creating repetitive execution conditions. Phi nodes manage loop indexes in LLVM. At each loop iteration start, they choose initial values. These values come from previous calculation results. In subsequent iterations, they choose indexes from back edges. 

% Phi nodes resovle conflicts between initial values at the loop iteration start and index values from back edges in subsequent iterations.
\begin{enumerate}
    \item If $g_j$ comes from a PHI node $p$, we recognize it as an index that changes during loop iterations. For such cases, we must check all incoming values $v_k$ from different predecessor basic blocks.
    The access is only considered safe when every incoming value $v_k$ stays within the permitted range (Lines {26-34}).

    \item If $g_j$ does not come from a PHI node {(Lines {35-37})}, i.e., $g_j$ is not relevant to loop iterations. In this case, we still need to confirm that $g_j$ remains within the safe range throughout execution.

\end{enumerate}

% In LLVM, natural loops that have a clear structure with a single entry (\texttt{Header}) and explicit back edges pointing to \texttt{Header}, can be identified and analyzed by LLVM.\footnote{\url{https://llvm.org/docs/LoopTerminology.html}} Due to natural loop semantics, $index$ remains constrained by the initial index value and compare instructions throughout loop execution. If the initial index value passes verification, even if the access dominates the compare in control flow, in subsequent iterations, the access instruction is still bounded by the compare instruction of the previous iteration. For example, in the loop shown in Listing \ref{lst:example} from Line 9 to 14, the index \texttt{j} for \texttt{buf[j]} stays strictly within [0, 20] with an initial index of \texttt{Entry = 0}, and the buffer size is exactly 20. This makes it safe to eliminate the sanitizer check.

% The same reasoning applies to loop-invariant memory accesses, such as \texttt{buf[0]}, where the offset can be verified as a constant safe value.  

% During implementation, if def-use analysis is blocked by PHI nodes in the loop header, the algorithm uses recursion to trace backward incoming values of PHI nodes and find index constraints.

\begin{algorithm}[tb]
\small
\caption{Removing redundant checks in loop}\label{alg:Removing Checks}  
\label{alg:identifying_asan_checks}
\SetKwProg{Fn}{Procedure}{}{}
\SetKwProg{Ag}{Algorithm}{}{}
\SetKwProg{Ed}{end}{}{}
\SetKwFunction{MustAlias}{must\_alias}
\SetKwFunction{Dominate}{dominate}
\SetKwFunction{PostDominate}{post\_dominate}
\SetKwFunction{ASan}{ASan}
\SetKwFunction{FindMemGroup}{is\_safe\_access}

\Fn{\FindMemGroup{$m_i, index, size$}}{
    % \tcp{Find the group $m$ belongs to}
    \If{$index$ is a constant}{
        \If{$\lnot (0 \le index < size) $}{
            \Return $false$\;
        }
        \Return $true$\;
    }
    \For{each $u_{q}$ $\in$ $index$.users()}{
        \If{$\lnot ( u_{q}$ is a compare instruction)}{
            \textbf{continue}\;
        }
        {$cmp \gets$ \texttt{dyncast\_to\_compare\_instruction}($u_{q}$)}\;
        $c$ $\gets$ \texttt{get\_constant}({$cmp$})\;
        \If{{$\lnot (cmp$ indicates "$index < c$") $\lor \lnot(0 \le c < size)$}}{
            \textbf{continue}\;
        }
       \If{$m_i$ dominates ${cmp}$ $\lor$ {$cmp$} dominates $m_i$}{
            % $is\_safe$ $\gets$ $true$\;
            \Return $true$\;
        } 
        
       %  $c$ $\gets$ \texttt{get\_constant}($u_{q}$)\;
       %  \If{$\lnot (u_{q}$ {indicates " $index < c$ "}) $\lor \lnot(0 \le c < size)$}{
       %      \textbf{continue}\;
       %  }
       % \If{$m_i$ dominates $u_{q} \lor u_{q}$ dominates $m_i$}{
       %      % $is\_safe$ $\gets$ $true$\;
       %      \Return $true$\;
       %  }   

        % \For{$m_j \in M_{grp}[m_i]$}{
        %     \If{$m = m_j \lor \MustAlias(m, m_j)$}{
        %         \Return $m_i$\;
        %     }
        %     \textbf{end}
        % }
        % \textbf{end}
    }
    % \textbf{end}
    
    % \Return \texttt{NULL}\;
    \Return $false$
}

\Ag{}{
    % \KwIn{A function $\mathcal{F}$}
    % \KwOut{Recurring ASan checks $\vec{R}_c = \{C_1, C_2, \dots, C_n\}$}
    % Initialization: $\vec{R}_c \leftarrow \emptyset$; $M_{grp} \leftarrow \texttt{dict}()$\;

    % \tcp{$M_{grp}$ is a dictionary where the key is a unique memory location and the value is a list of memory accesses to that location}
    % \ForEach{memory access $m_i \in \mathcal{F}$}{
    %     $m_j \leftarrow \FindMemGroup(M_{grp}, m_i)$\;
    %     \If{$m_j = \texttt{NULL}$}{
    %         $M_{grp}[m_i] \leftarrow \texttt{list}()$\;
    %         $m_j \leftarrow m_i$\;
    %     }
    %     \textbf{end}
        
    %     $M_{grp}[m_j].\texttt{add}(m_i)$\;
    % }
    % \textbf{end}

    % \For{$m \in M_{grp}$}{
    %     \For{$m_i \in M_{grp}[m]$}{
    %         \For{$m_j \in M_{grp}[m]$}{
    %             \If{$m_i \ne m_j \land \Dominate(m_i, m_j) \land sizeof(m_i) \ge sizeof(m_j)$}{
    %                 $\vec{R}_c.\texttt{add}(\ASan(m_j))$\;
    %                 $M_{grp}[m].\texttt{remove}(m_j)$\;
    %             }
    %             \textbf{end}
    %         }
    %         \textbf{end}
    %     }
    %     \textbf{end}
        
    %     \For{$m_i \in M_{grp}[m]$}{
    %         \For{$m_j \in M_{grp}[m]$}{
    %             \If{$m_i \ne m_j \land \PostDominate(m_i, m_j) \land sizeof(m_i) \ge sizeof(m_j)$}{
    %                 $\vec{R}_c.\texttt{add}(\ASan(m_j))$\;
    %                 $M_{grp}[m].\texttt{remove}(m_j)$\;
    %             }
    %             \textbf{end}
    %         }
    %         \textbf{end}
    %     }
    %     \textbf{end}
    % }
    % \textbf{end}

    \KwIn{An LLVM IR function $\mathcal{F}$}
    \KwOut{A safe memory access instruction set $\mathcal{R}$}
    $\mathcal{R} \gets \phi$\;
    $\mathcal{M}$ $\gets$ \texttt{find\_all\_interesting\_memory\_access}($\mathcal{F}$)\; 

    \For{each memory access instruction $m_i$ $\in$ $\mathcal{M}$}{
        % $depth$ $\gets$ \texttt{get\_loop\_depth}($m_i$)\;
        \If{\texttt{get\_loop\_depth}($m_i$) $ \neq 1$}{
            \textbf{continue}\;
        }
        % $gep$ $\gets$ \texttt{find\_}$getelementptr$\texttt{\_instruction}($m_i$)\;
        $gep$ $\gets$ \texttt{find\_getelementptr\_instruction}($m_i$)\;
        $size$ $\gets$ \texttt{get\_access\_object\_size}($gep$)\;
        % $safe$ $\gets$ $true$\;
        $all\_g_j\_is\_safe \gets true$ \;
        \For{each $g_j$ $\in$ $gep$.indexes()}{
            % \If{$g_j$ is a constant}{
            %     \If{$g_j > size$}{
            %         % $is\_safe \gets false$\;
            %         \textbf{break}\;
            %     }
            %     \textbf{continue}\;
            % }
            \If{$g_j$ comes from phi node p}{
                $all\_v_k\_is\_safe \gets true$ \;
                \For{each $v_k$ $\in$ p.incoming\_values()}{

                    \If{$\lnot$ \FindMemGroup($ m_{i}, v_{k}, size$)}{
                        $all\_v_k\_is\_safe \gets false$ \;
                        \textbf{break}\;
                    }
                }
                \If{{$\lnot all\_v_k\_is\_safe$}}{
                    % $\mathcal{R} \cup \{m_i\}$\;
                    $all\_g_j\_is\_safe \gets false$ \;
                    \textbf{break}\;
                }
            }
            % \tcp{command}
            \ElseIf{$\lnot$\FindMemGroup($ m_{i}, g_{j}, size$)}{
                % $is\_safe \gets false$\;
                % \For{each user $u_k$ of $g_i$.users()}{
                %     \If{$u_k$ is not a compare instruction}{
                %         \textbf{continue}\;
                %     }
                %     $c \gets$ \texttt{get\_constant}($u_k$)\;
                %     \If{$u_k$ describes ($g_j <= c$)}{
                %         \textbf{continue}\;
                %     }
                %     \If{($m_i$ dominates $u_k$ \texttt{||} $u_k$ dominates $m_i$) \texttt{ \&\& } $size >= c$}{
                %         % $is\_safe$ $\gets$ $true$\;
                %         $\mathcal{R} \cup \{m_i\}$\;
                %         \textbf{break}\;
                %     }
                % }
                $all\_g_j\_is\_safe \gets false$ \;
                \textbf{break}\;
            }
        }
        \If{{$all\_g_j\_is\_safe$}}{
            $\mathcal{R} \cup \{m_i\}$\;
        }
    }
% \end{algorithm}
    \Return{$\mathcal{R}$}\;
}

\end{algorithm}

% Due to compilation performance trade-offs, Algorithm 1 currently optimizes only single-level loops. Despite this limitation, our evaluation on gcc-loop benchmarks shows our optimizer with Algorithm \ref{alg:Removing Checks} eliminates 50\% more checks in loop and  12\% more checks in the whole program than ASan\mbox{-}\mbox{-}. This translates to a 41\% reduction in runtime overhead compared to ASan\mbox{-}\mbox{-}.

Due to the trade-off between compilation performance and runtime performance, Algorithm \ref{alg:Removing Checks}  only removes redundant checks in single-level loops {(Line 20)}. Despite this limitation, our evaluation on a loop-intensive benchmark {gcc-loops}\footnote{\url{https://github.com/llvm/llvm-test-suite/blob/main/SingleSource/UnitTests/Vectorizer/gcc-loops.cpp}} shows our optimizer with Algorithm \ref{alg:Removing Checks} eliminates 18\% of checks in loops. When the two-stage check is disabled, \textsc{Tech-ASan} using only the optimizer introduces 102\% runtime overhead on {gcc-loops}, compared to 144\% for ASan\mbox{-}\mbox{-}. This demonstrates the efficiency of Algorithm \ref{alg:Removing Checks}.

\begin{sloppypar}

For checks in loops that access contiguous memory regions and can not be optimized by Algorithm \ref{alg:Removing Checks}, we explore an alternative optimization approach. We first leverage LLVM’s built-in Scalar Evolution (SCEV) analysis to infer the start addresses and the sizes of memory accesses within the loops. 
Then we insert ASan’s optimized region-checking function \texttt{\_\_asan\_region\_is\_poisoned()} at the exit block of the loop.
This technique demonstrates significant speedups in benchmarks where large arrays are accessed iteratively. However, in SPEC CPU2006 \cite{SPECCPU2006}, measurable performance gains are observed only in a small subset of programs due to two key limitations: First,
\texttt{\_\_asan\_region\_is\_poisoned()} incurs expensive calling costs, which outweighs its benefits when the accessed contiguous memory regions are short.
Second, for loops with variable iterations, the length of continuous accesses cannot be statically determined at compile time. Since many loops in SPEC CPU2006 operate on short memory regions, this approach introduces overhead rather than optimization.
Given these trade-offs, we remove this optimization in our final implementation.
\end{sloppypar}

In addition, we successfully integrate the other SOTA redundant check elimination techniques into the optimizer of \textsc{Tech-ASan}, including \textit{removing recurring checks} and \textit{optimizing neighbor checks} \cite{ASAN--2022,FloatZone2023}. First, removing recurring checks identifies checks sharing the same memory location and access size in each function using LLVM built-in alias-analysis, if they either all happen or all do not happen, we only need to retain a single check. Second, optimizing neighbor checks is to merge or remove adjacent checks. When checks either both happen or both do not happen, there are two situations that can be optimized: (1) If two memory accesses can fall into an 8-byte region, their checks can be merged into a single check.
(2) Given three memory accesses, i.e., (\texttt{addr1}, \texttt{size1}), (\texttt{addr2}, \texttt{size2}), and (\texttt{addr3}, \texttt{size3}), where \texttt{addr1 < addr2 < addr3}. \textsc{Tech-ASan}'s check for the second access, i.e., (\texttt{addr2}, \texttt{size2}), can be safely eliminated if \texttt{addr3 - addr1 < MinRdSz} and  \texttt{addr2 + size2 $\le$ addr3 + size3}, where \texttt{MinRdSz} is the minimal size of a redzone.

\begin{table*}[t]
\small
  \centering
  \caption{Detection capability on the adjusted Juliet Test Suite with buggy and non-buggy testcases}
  \label{tab:JTS}%
    \begin{tabular}{llrrrrr|rrrrr}
    \toprule
    \multirow{2}{*}{\textbf{CWE \& Type}} & \multirow{2}{*}{\textbf{Total}} & \multicolumn{5}{c|}{\textbf{Buggy testcases}} & \multicolumn{5}{c}{\textbf{Non-buggy testcases}} \\
     &  & {\textsc{Tech-ASan}} & {ASan} & {ASan\mbox{-}\mbox{-}} & {GiantSan} & {FloatZone} & {\textsc{Tech-ASan}} & {ASan}  & {ASan\mbox{-}\mbox{-}} & {GiantSan} & {FloatZone} \\
    \midrule
    121: stack-buffer-overflow & 2956  & 2956  & 2948  & 2948  & 2611  & 2822  & 2956  & 2956  & 2956  & 2956  & 2908 \\
    % \midrule
    122: heap-buffer-overflow & 3438  & 3438  & 3390  & 3390  & 3008  & 3337  & 3438  & 3438  & 3438  & 3438  & 3366 \\
    % \midrule
    124: buffer-underwrite & 1024  & 1024  & 1024  & 1024  & 1021  & 976   & 1024  & 1024  & 1024  & 1024  & 1024 \\
    % \midrule
    126: buffer-overread & 672   & 672   & 672   & 672   & 493   & 672   & 672   & 672   & 672   & 672   & 672 \\
    % \midrule
    415: double-free & 818   & 818   & 818   & 818   & 818   & 818   & 818   & 818   & 818   & 818   & 818 \\
    % \midrule
    416: use-after-free & 393   & 393   & 393   & 393   & 381   & 393   & 393   & 393   & 393   & 393   & 393 \\
    \midrule
    \textbf{Total} & \textbf{9301} & \textbf{9301} & \textbf{9245} & \textbf{9245} & \textbf{8332} & \textbf{9018} & \textbf{9301} & \textbf{9301} & \textbf{9301} & \textbf{9301} & \textbf{9181} \\
    \bottomrule
    \end{tabular}%
  % \label{tab:addlabel}%
\end{table*}%

\subsection{Implementation}
We have implemented \textsc{Tech-ASan} on top of the ASan's infrastructure in the LLVM compiler \cite{LLVM2004}.
Compared to the native ASan, our implementation introduces a total of 2.4K additional lines of code. The use of \textsc{Tech-ASan} is identical to ASan because \textsc{Tech-ASan} is implemented to share the same assumptions, requirements, and interfaces with ASan. Our design is fully generalizable to other compilers. We have not added any architecture-specific code, allowing \textsc{Tech-ASan} to run on any machine with instruction set architectures supported by LLVM's backend. Therefore, \textsc{Tech-ASan} maintains ASan's usability.

As demonstrated in our experiments, \textsc{Tech-ASan} does not compromise the detection capabilities of ASan while reducing the size of the binaries compiled with ASan. The additional compilation time introduced by \textsc{Tech-ASan} is also within an acceptable range.

\section{Evaluation}\label{sec:Evaluation}

We experimentally evaluate \textsc{Tech-ASan} on the following research questions (RQs):
\begin{itemize}
    \item \textbf{RQ1:} Can \textsc{Tech-ASan} maintain the detection capability of ASan? 
    % (\S\ref{sec:Detection Capability})
    \item \textbf{RQ2:} Can \textsc{Tech-ASan} reduce the runtime overhead of ASan?
    % (\S\ref{sec:Runtime overhead})
    \item \textbf{RQ3:} How does \textsc{Tech-ASan} impact compilation time and binary size compared to ASan? 
    % (\S\ref{sec:Binary Size}) 
    % (\S\ref{sec:Compilation Efficiency})
    % \item \textbf{RQ4:} Can \textsc{Tech-ASan} reduce the size of compiled binaries compared to ASan? (\S\ref{sec:Binary Size})
\end{itemize}
By inheriting ASan's shadow memory model, \textsc{Tech-ASan} incurs no additional memory overhead, therefore, comparative memory analysis is not needed. When evaluating the detection capability of \textsc{Tech-ASan}, all compiler optimizations are disabled to prevent vulnerabilities from being suppressed by optimization passes. For other measurements, the optimization option \texttt{Og} is enabled by default. Unless otherwise specified, all the experiments are conducted on an x86-64 server running Ubuntu 18.04, equipped with an AMD EPYC 7702 192-core 2.0GHz CPU and 128GB RAM.

\begin{table*}[tbp]
\small
  \centering
  \caption{Detection capability for real-world memory safety violations from CVE. \ding{51}, \ding{53}, and - represent that a violation is detected, a violation is not detected, the program can not be instrumented or executed normally, respectively.} 
    \begin{tabular}{llrllcccc}
    \toprule
    % \hline
    \textbf{Program} & \textbf{Version} & \textbf{LoC (k)} & \textbf{Vulnerabilities} & \textbf{Vulnerability Type} & \textbf{\textsc{Tech-ASan}} & \textbf{ASan} & \textbf{ASan\mbox{-}\mbox{-}} & \textbf{GiantSan} \\
    % \hline
    % \hline
    \midrule
    binutils & 2.15  & 1456.9 & CVE-2006-2362 & stack-buffer-overflow & \ding{51}     & \ding{51}     & \ding{51}     & \ding{51} \\
    % \hline
    binutils & 2.29  & 3888.6 & CVE-2018-9138 & stack-overflow & \ding{51}     & \ding{51}     & \ding{51}     & \ding{51} \\
    % \hline
    fcron & 3.0.0 & 35.7 & CVE-2006-0539 & heap-buffer-overflow & \ding{51}     & \ding{51}     & \ding{51}     & \ding{51} \\
    % \hline
    fig2dev & 3.2.7b & 49.7 & CVE-2020-21675 & stack-buffer-overflow & \ding{51}     & \ding{51}     & \ding{51}     & - \\
    % \hline
    {GraphicsMagick} & 1.3.26 & 476.8 & CVE-2017-12937 & heap-buffer-overflow & \ding{51}     & \ding{51}     & \ding{51}     & \ding{51} \\
    % \hline
    GoHttp &      & 0.5   & CVE-2019-12160 & heap-use-after-free & \ding{51}     & \ding{51}     & \ding{51}     & \ding{51} \\
    % \hline
    libpng & 1.6.37 & 94.1 & CVE-2021-4214 & heap-buffer-overflow & \ding{51}     & \ding{51}     & \ding{51}     & \ding{53} \\
    % \hline
    libtiff & 3.8.0 & 113.5 & CVE-2006-2025 & Integer-overflow & \ding{51}     & \ding{51}     & \ding{51}     & \ding{51} \\
    % \hline
    libtiff & 3.8.2 & 112.4 & CVE-2009-2285 & heap-buffer-overflow & \ding{51}     & \ding{51}     & \ding{51}     & \ding{51} \\
    % \hline
    \multirow{2}{*}{libtiff} & \multirow{2}{*}{4.0.1} & \multirow{2}{*}{126.8} & CVE-2013-4243 & heap-buffer-overflow & \ding{51}     & \ding{51}     & \ding{51}     & \ding{53} \\
% \cline{4-9}
&       &       & CVE-2015-8668 & heap-buffer-overflow & \ding{51}     & \ding{51}     & \ding{51}     & \ding{51} \\
    % \hline
    libzip & 1.2.0 & 48.1 & CVE-2017-12858 & heap-use-after-free & \ding{51}     & \ding{51}     & \ding{51}     & \ding{51} \\
    % \hline
    lua   & 5.4.3 & 31.7 & CVE-2021-44964 & heap-use-after-free & \ding{51}     & \ding{51}     & \ding{51}     & \ding{51} \\
    % \hline
    \multirow{3}{*}{mp3gain} & \multirow{3}{*}{1.5.2} & \multirow{3}{*}{9.1} & CVE-2017-14407 & stack-buffer-overflow & \ding{51}     & \ding{51}     & \ding{51}     & \ding{53} \\
% \cline{4-9}  
&       &       & CVE-2017-14408 & stack-buffer-overflow & \ding{51}     & \ding{51}     & \ding{51}     & \ding{51} \\
% \cline{4-9}
&       &       & CVE-2017-14409 & global-buffer-overflow & \ding{51}     & \ding{51}     & \ding{51}     & \ding{53} \\
    % \hline
    mxml  & 2.12  & 27.8 & CVE-2018-20004 & stack-buffer-overflow & \ding{51}     & \ding{51}     & \ding{51}     & \ding{51} \\
    % \hline
    nasm  & 2.15.04rc3 & 155.0 & CVE-2020-24978 & double-free & \ding{51}     & \ding{51}     & \ding{51}     & \ding{51} \\
    % \hline
    python & 3.1.5 & 692.0 & CVE-2014-1912 & heap-use-after-free & \ding{51}     & \ding{51}     & \ding{51}     & \ding{51} \\
    % \hline
    yasm  & 1.3.0 & 164.9 & CVE-2021-33468 & heap-use-after-free & \ding{51}     & \ding{51}     & \ding{51}     & \ding{51} \\
    % \hline
    \midrule
    % & & & & & & & & \\
    \textbf{Total} &  & \textbf{7483.6} & \textbf{20} &  & \textbf{20} & \textbf{20} & \textbf{20} & \textbf{15} \\

    \bottomrule
    % \hline
    \end{tabular}%
  % \label{tab:addlabel}%
  \label{tab:CVE}
\end{table*}%

\subsection{RQ1: Detection Capability}\label{sec:Detection Capability}
To measure the detection capability of \textsc{Tech-ASan}, we conduct two experiments. The first experiment is conducted on the latest version of the Juliet Test Suite (version 1.3) \cite{black2018juliet}, where all CWEs have both buggy and non-buggy testcases, which are used to test the false negatives and false positives of the sanitizers, respectively. However, there are 4 types of testcases that are not suitable for evaluation: 
\begin{enumerate}
    \item Testcases that wait for an external signal, e.g., sockets. We just remove them to avoid waiting infinitely.
    \item Testcases with violations that are triggered depend on a random number. We change them to the non-random versions for a deterministic result.
    \item Testcases that
    % with names containing \texttt{CWE170\_char\_*} 
    print a string without the terminating character, which  produces random overflow. We remove them for a deterministic result. 
    \item Testcases with violations that are only triggered on 32-bit systems but are not triggered on 64-bit systems.
\end{enumerate}
% Since Juliet Test Suite contains testcases that wait for an external signal (e.g., sockets), and some testcases include a randomized version (triggered with probability).
After adjustments, the number of remaining testcases in each CWE type is still much more than that used in recent works \cite{ASAN--2022, giantsan2024,FloatZone2023} for the more comprehensive evaluation. The adjusted Juliet Test Suite has been published in the community for future science.\footnote{\url{https://github.com/Hufffman/Adjusted-Juliet-Test-Suite}}
% Note that ASan can not pass all testcases since that several memory safety violations can not be triggered. For example, in CWE122, a heap-buffer-over-flow testcase looks like:
% \begin{verbatim}
% double * data = NULL;
% data = (double *)malloc(sizeof(data));
% if (data == NULL) {
%     exit(-1);
% }
% *data = 1.7E300; // buffer-over-flow on 32-bit machine
% \end{verbatim}
% This violation can be triggered on 32-bit machine, because the size of the heap buffer pointed by \texttt{data} is 4 byte and \texttt{*data} loads a 8-byte value. But it can be not triggered on our 64-bit machine because \texttt{sizeof(data)} is 8.


For comparison, we further run recent available methods ASan\mbox{-}\mbox{-} \cite{ASAN--2022}, GiantSan \cite{giantsan2024}, and FloatZone \cite{FloatZone2023} on the Juliet Test Suite. Note that  FloatZone is executed on a 64-bit server equipped with an Intel CPU since its public version can be run on Intel CPUs only.
Table \ref{tab:JTS} shows that only \textsc{Tech-ASan} has no false positive and false negative issues. 
On buggy testcases, ASan, ASan\mbox{-}\mbox{-}, GiantSan, and FloatZone can not detect 56 memory safety violations that occur in \texttt{wcscpy()}, which remains unsolved even in the latest version of ASan. GiantSan generates an additional 913 false negatives, due to the insufficient check on the stack arrays, failing to report an error when an out-of-bounds write occurs to a stack array within a loop.
FloatZone generates 283 false negatives due to the following reasons:
(1) In testcases containing both partial and full out-of-bounds (OOB) accesses, partial OOB operations corrupt the redzone headers. This corruption invalidates FloatZone's check logic that depends on a complete redzone, consequently failing to detect subsequent full OOB accesses. (2) In some {testcases}, FloatZone fails to insert redzones for memory objects when allocating via C standard library functions, e.g., \texttt{malloc()}.

On non-buggy testcases, \textsc{Tech-ASan}, ASan, ASan\mbox{-}\mbox{-}, and GiantSan have no false positive issue. However, FloatZone produces 120 false positives in buffer overflow detection due to erroneous reports when programs access residual redzones in reused stack memory regions.

% 备注：
% 122中,并非漏报。在64位系统上,sizeof(void*)的大小和double*的大小是一样的。
% void CWE122_Heap_Based_Buffer_Overflow__sizeof_double_41_bad()
% {
%     double * data;
%     /* Initialize data */
%     data = NULL;
%     /* INCIDENTAL: CWE-467 (Use of sizeof() on a pointer type) */
%     /* FLAW: Using sizeof the pointer and not the data type in malloc() */
%     data = (double *)malloc(sizeof(data));
%     if (data == NULL) {exit(-1);}
%     *data = 1.7E300;
%     CWE122_Heap_Based_Buffer_Overflow__sizeof_double_41_badSink(data);
% }

In the second experiment, we select real-world memory-related memory safety violations from common vulnerabilities and exposures (CVEs) \cite{vulnerabilities2005common}, including 20 CVEs with known proof-of-concepts (PoCs) from 14 projects written in C/C++. Since Table \ref{tab:JTS} shows that FloatZone may produce false positives, we exclude it to avoid uncertainty in confirming whether a vulnerability is actually detected. Table \ref{tab:CVE} shows that ASan, ASan\mbox{-}\mbox{-}, and \textsc{Tech-ASan} detect all the CVEs. However, GiantSan fails to detect four known CVEs, and triggers an assertion failure, leading to a program crash during the reproduction of an additional CVE.

In summary, \textsc{Tech-ASan} preserves ASan's detection capability while accelerating and eliminating sanitizer checks.

\subsection{RQ2: Runtime Overhead}\label{sec:Runtime overhead}

\begin{table*}[tb]
\small
  \centering
  \caption{Runtime overhead on SPEC CPU2006 Benchmark (seconds). Ratio $R$ is computed as the execution time of sanitizer-enabled binaries divided by that of non-instrumented vanilla clang binaries. The "-" indicates abnormal program termination during testing.
  }  \label{tab:spec_runtime}%
    \begin{tabular}{l|rrcrcrcrc|rc}
    \toprule
    \multicolumn{1}{c|}{\multirow{2}{*}{\textbf{Programs}}} & \multicolumn{9}{c|}{\textbf{Performance Study}}                        & \multicolumn{2}{c}{\textbf{Ablation Study}} \\
         & Vanilla Clang & \textsc{Tech-ASan} & \textit{R}     & ASan  &\textit{R}    & ASan\mbox{-}\mbox{-} &\textit{R}    & {GiantSan} &\textit{R}    & $\textsc{Tech-ASan}^{\text{ac}}$ & \textit{R} \\
         \midrule
    % \toprule
    400.perlbench & {262} & 1415  & 540.08\% & {1600} & 610.69\% & {1390} & 530.53\% & {1440} & 549.62\% & {1390} & 530.53\% \\
    % \midrule
    401.bzip2 & {393} & 502   & 127.61\% & {623} & 158.52\% & {588} & 149.62\% & {552} & 140.46\% & {524} & 133.21\% \\
    % \midrule
    403.gcc & {242} & 1145  & 473.14\% & {959} & 396.28\% & {929} & 383.88\% & {857} & 354.13\% & {1180} & 487.60\% \\
    % \midrule
    429.mcf & {390} & 553   & 141.67\% & {738} & 189.23\% & {725} & 185.90\% & {662} & 169.74\% & {568} & 145.51\% \\
    % \midrule
    445.gobmk & {358} & 493   & 137.71\% & {625} & 174.58\% & {547} & 152.79\% & {664} & 185.47\% & {497} & 138.83\% \\
    % \midrule
    456.hmmer & {348} & 429   & 123.28\% & {646} & 185.63\% & {568} & 163.22\% & {400} & 114.94\% & {475} & 136.35\% \\
    % \midrule
    458.sjeng & {489} & 707   & 144.48\% & {898} & 183.64\% & {742} & 151.74\% & -     & -     & {721} & 147.34\% \\
    % \midrule
    462.libquantum & {513} & 622   & 121.25\% & {672} & 130.99\% & {682} & 132.94\% & {603} & 117.54\% & {631} & 122.90\% \\
    % \midrule
    473.astar & {448} & 559   & 124.67\% & {664} & 148.21\% & {595} & 132.81\% & {487} & 108.71\% & {569} & 127.01\% \\
    % \midrule
    483.xalancbmk & {823} & 1525  & 185.30\% & {1740} & 211.42\% & {1530} & 185.91\% & {1380} & 167.68\% & {1545} & 187.73\% \\
    % \midrule
    433.milc & {537} & 807   & 150.19\% & {1125} & 209.50\% & {976} & 181.75\% & {798} & 148.60\% & {835} & 155.40\% \\
    % \midrule
    444.namd & {320} & 379   & 118.44\% & {518} & 161.88\% & {530} & 165.63\% & {374} & 116.88\% & {394} & 122.97\% \\
    % \midrule
    447.dealII & {1100} & 1520  & 138.18\% & {1770} & 160.91\% & {1720} & 156.36\% & {1520} & 138.18\% & {1540} & 140.00\% \\
    % \midrule
    450.soplex & {318} & 466   & 146.38\% & {517} & 162.58\% & {491} & 154.40\% & {457} & 143.71\% & {479} & 150.63\% \\
    % \midrule
    453.povray & {179} & 451   & 251.96\% & {544} & 303.91\% & {453} & 253.07\% & {452} & 252.51\% & {452} & 252.23\% \\
    % \midrule
    470.lbm & {242} & 264   & 109.09\% & {322} & 133.06\% & {289} & 119.42\% & {257} & 106.20\% & {279} & 115.29\% \\
    % \midrule
    482.sphinx3 & {364} & 521   & 143.13\% & {616} & 169.23\% & {615} & 168.96\% & {445} & 122.25\% & {508} & 139.56\% \\
    \midrule
    \textbf{geomean} &       &       & \textbf{164.00\%} &       & \textbf{197.70\%} &       & \textbf{181.89\%} &       & \textbf{162.64\%} &       & \textbf{167.91\%} \\
\bottomrule
% \cmidrule{2-12} 
\end{tabular}%
\end{table*}%
Following the recent studies \cite{FloatZone2023,ASAN--2022,Catamaran2023,OPTISAN2024}, we use the classic version of the industry-standard CPU-intensive runtime benchmark suite, SPEC CPU2006 \cite{SPECCPU2006}, to evaluate the performance improvement of \textsc{Tech-ASan} thoroughly. Since some of the programs in SPEC CPU2006 contain memory safety violations, 
all programs are compiled with \texttt{-fsanitize-recover=address} and run after setting the environment variable \texttt{ASAN\_OPTIONS} to \texttt{"halt\_on\_error=0"}, which ensures that sanitizers continue execution after detecting a violation, rather than halting the program. By analyzing the logs, we confirm that \textsc{Tech-ASan} correctly detects violations without any false positives. To mitigate accidental errors, we calculate the median of the 10 running times.

Table \ref{tab:spec_runtime} presents the runtime performance of the SPEC CPU2006 benchmark. \textsc{Tech-ASan}, ASan, ASan\mbox{-}\mbox{-}, and GiantSan introduce {64.0\%, 97.7\%, 81.89\%, and 62.64\% runtime overhead compared to the vanilla clang.} \textsc{Tech-ASan} reduces ASan's runtime overhead by 33.7\%, and outperforms ASan\mbox{-}\mbox{-} by 17.89\%, demonstrating the effectiveness of \textsc{Tech-ASan}'s two-stage check and redundant check elimination techniques. GiantSan introduces the lowest runtime overhead, due to its operation-level instrumentation and history caching mechanism. Although \textsc{Tech-ASan} has 1.36\% higher overhead than GiantSan, it remains within an acceptable range due to its detection capability. Specifically, \textsc{Tech-ASan} outperforms GiantSan on 429.mcf and 445.gobmk, because GiantSan's operation-level check merging provides less effectiveness due to non-continuous memory access such as tree and graph. In addition, \textsc{Tech-ASan} exhibits higher runtime overhead than ASan on 403.gcc, resulting from the frequent memory allocation and free, which increases the runtime overhead of magic value injection.

An ablation study is conducted to further clarify the runtime benefits brought by \textsc{Tech-ASan}'s check acceleration technique. Let $\textsc{Tech-ASan}^\text{ac}$ represents that \textsc{Tech-ASan} only enables check acceleration. Table \ref{tab:spec_runtime} presents that $\textsc{Tech-ASan}^\text{ac}$ reduces ASan's overhead by 29.79\%, which demonstrates that our design of two-stage check is effective to speed up sanitizer checks and reduce the runtime overhead. \textsc{Tech-ASan} achieves 3.91\% lower overhead than $\textsc{Tech-ASan}^\text{ac}$ due to the optimizer. Furthermore, as the overhead of individual sanitizer checks is significantly reduced by the two-stage check mechanism, the efficacy of the optimizer becomes less pronounced in this context.

\subsection{RQ3: Usability Analysis}
\label{sec:Binary Size}

This section evaluates  \textsc{Tech-ASan}'s binary size expansion and compilation time overhead compared to ASan.

\begin{figure*}[tb]
  \centering
  \includegraphics[width=\linewidth]{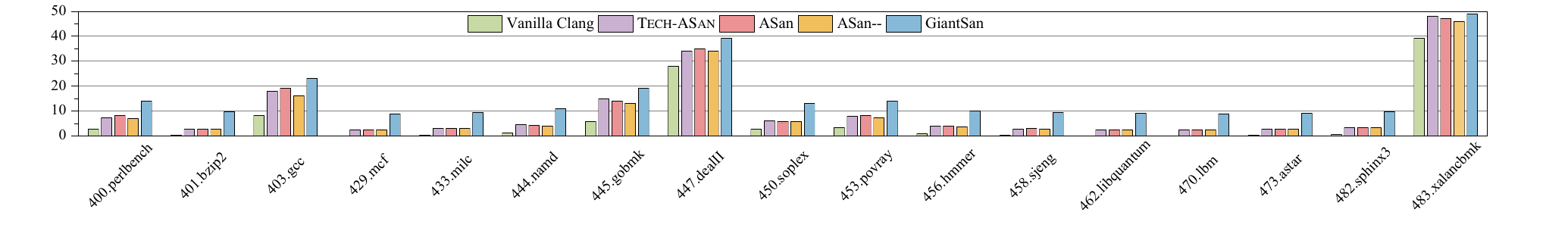}
  \caption{The binary sizes of programs in SPEC CPU2006 benchmark (MB)}
  \label{fig:binary-size}
\end{figure*}

The binary size is a critical factor in the deployment of memory safety tools, especially in environments with limited storage capacity, such as embedded systems, Internet of Things (IoT) devices, and mobile applications. Figure \ref{fig:binary-size} presents the binary sizes of programs in the SPEC CPU2006 benchmark. The average sizes of binary code compiled with \textsc{Tech-ASan}, ASan, ASan\mbox{-}\mbox{-}, and GiantSan are 11.64$\times$, 11.78$\times$, 11.52$\times$,  and 38.94$\times$ those of vanilla clang, respectively. ASan\mbox{-}\mbox{-} eliminates partial redundant checks and introduces minimal additional code for its "loop check optimization" feature, thereby reducing binary size, which is consistent with observations in \cite{ASAN--2022}.
Compared to ASan and ASan\mbox{-}\mbox{-}, although \textsc{Tech-ASan} introduces additional code for redzone management and replaces ASan's native check logic with a two-stage check logic, it still reduces the average binary size by 14.79\% comparable to ASan. This improvement is due to two reasons: First, \textsc{Tech-ASan} implements second stage checks through function calls rather than inline instruction sequences. Second, \textsc{Tech-ASan}'s optimizer incorporates a novel algorithm for eliminating redundant checks within loops. GiantSan's implementation of complicated operation-level protection and shadow memory encoding mechanisms significantly increases instrumented code size.

\begin{table}[tb]
\small
  \centering
  \caption{Compile-time overhead on SPEC CPU2006, each number is the ratio compared to the native compilation. }
    \begin{tabular}{lrrrr}
    \toprule
    \textbf{Programs} & \textbf{\textsc{Tech-ASan}} & \textbf{ASan}   & \textbf{ASan\mbox{-}\mbox{-}} & \textbf{GiantSan} \\
    \midrule
    400.perlbench & 286.11\% & 159.15\%   & 287.12\% & 657.81\% \\
    401.bzip2 & 157.87\% & 130.64\%   & 135.96\% & 3640.85\%\\
    403.gcc & 169.70\% &  162.17\%   & 189.15\%  & 318.48\%\\
    429.mcf & 105.43\% & 103.70\%  & 107.16\% & 118.27\%  \\
    433.milc & 113.63\% & 114.24\%   & 114.09\% & 133.69\%\\
    444.namd & 162.85\% & 136.83\%  & 182.06\% & 224.34\%  \\
    445.gobmk & 133.10\% & 148.38\%   & 152.37\% & 227.93\% \\
    447.dealII & 120.48\% & 118.04\%  & 121.78\% & 205.34\%  \\
    450.soplex & 112.97\% & 111.77\%   & 117.97\% & 172.03\%\\
    453.povray & 125.92\% & 134.08\%  & 133.41\% & 206.49\%  \\
    456.hmmer & 128.27\%  & 132.82\%   & 134.81\% & 177.72\% \\
    458.sjeng & 108.56\% & 122.46\%  & 114.62\% & 142.07\%  \\
    462.libquantum & 110.26\% & 104.77\%   & 105.25\% & 127.92\% \\
    470.lbm & 112.67\% & 101.38\%  & 108.54\% & 125.07\%  \\
    473.astar & 108.05\% & 109.32\%   & 113.35\% & 125.64\% \\
    482.sphinx3 & 112.75\% & 117.00\%  & 121.53\% & 174.79\%  \\
    483.xalancbmk & 112.95\% & 114.09\%   & 115.89\% & 178.31\% \\
    \midrule
    \textbf{geomean} & \textbf{129.52\%}  & \textbf{123.50\%}  & \textbf{133.46\%} & \textbf{220.93\%} \\
    \bottomrule
    \end{tabular}%
  \label{tab:compile time}%
\end{table}%

Compilation efficiency is also a critical factor in the adoption of memory safety tools, especially in large-scale projects where frequent recompilation is required. To ensure statistically accurate compilation time measurements, we employ single-threaded compilation throughout the experiments. Table \ref{tab:compile time} presents the compile time of programs in the SPEC CPU2006 benchmark.
\textsc{Tech-ASan}, ASan, ASan\mbox{-}\mbox{-}, and GiantSan increase compile time by 29.52\%, 23.50\%, 33.46\%, and 120.93\% compared to vanilla clang, respectively. Compared to vanilla clang, the additional compilation time mainly comes from two sources: instrumentation and analysis. Compared to ASan, ASan\mbox{-}\mbox{-} employs extensive static analysis for which checks can be safely eliminated, consequently introducing additional compilation overhead. In contrast, \textsc{Tech-ASan} utilizes a more lightweight loop analysis approach, resulting in a 3.94\% reduction in compilation time relative to ASan\mbox{-}\mbox{-}. GiantSan incurs significant analysis overhead to merge multiple instruction-level instrumentations into an operation-level instrumentation, resulting in a longer compilation time compared to others. Notably, check elimination can reduce compilation time in certain cases. For instance, both \textsc{Tech-ASan} and ASan\mbox{-}\mbox{-} compile 458.sjeng faster than ASan. This occurs because the analysis overhead is offset by reduced instrumentation costs and decreased pressure on the compiler backend. Overall, \textsc{Tech-ASan} introduces only 29.52\% compile-time overhead on SPEC CPU2006, which represents an acceptable cost for large-scale software compilation.

\section{Related Work}\label{sec:related works}

This section reviews the related work on mitigating and detecting memory safety violations in C/C++ programs at runtime. Existing solutions can be categorized into \textit{location-based} methods \cite{ASAN--2022, ASAN2012, FloatZone2023, Memcheck2007,giantsan2024,doubletake2016,dr.memory2011} and \textit{pointer-based}  methods \cite{PACMem2022,UAFSan2021,hwasan2018,HWASanIO2023,dangzero2022,oscar2017,softbound2009,cets2010,EffectiveSan2018,Sulong2018}.

\subsection{Location-based Methods}
Location-based methods model the memory with a focus on memory bytes by recording which byte is addressable at runtime.
In general, these methods use canary values or shadow bytes to mark allocated memory as addressable and unallocated or freed memory as unaddressable. Location-based methods are widely deployed for their high compatibility.

Memcheck \cite{Memcheck2007} and Dr. Memory \cite{dr.memory2011} utilize shadow memory to track the state of each memory byte. Both tools incur more than a 10× runtime overhead due to the use of a dynamic binary instrumentation framework \cite{valgrind2007,DynamoRIO2004}. ASan \cite{ASAN2012}, the SOTA location-based method, instruments the tested programs at compile-time and leverages a compact state encoding in shadow memory to record which byte is addressable at runtime. DoubleTake \cite{doubletake2016} uses canary values to mark unaddressable memory locations and divides program execution into multiple epochs. It checks whether the canary value has been modified at the end of each epoch to determine whether a memory safety violation has occurred. However, DoubleTake does not provide adequate protection for freed memory objects, and the canary-based mechanism can only detect write vulnerabilities, but not read vulnerabilities. ASan\mbox{-}\mbox{-} \cite{ASAN--2022} reduces ASan's redundant checks at compile time through static analysis, provided safety is guaranteed. FloatZone \cite{FloatZone2023} leverages the floating-point unit (FPU) in the CPU to speed up sanitizer checks, but inevitably introduces false positives. GiantSan \cite{giantsan2024} proposes a shadow encoding with segment folding to increase the protection density and introduces operation-level protection to accelerate the sanitizer checks.

% Recently, Catamaran \cite{Catamaran2023} is proposed to accelerate both pointer-based methods and location-based methods 

However, a prior study \cite{UAFSan2021} shows that location-based methods may miss use-after-free vulnerabilities. To mitigate this, a commonly adopted solution is to quarantine freed heap objects in a queue. Once the queue is full, the oldest objects are popped and reallocated to the program, which can still lead to false negatives. Fortunately, few reports related to false negatives exist in practice due to the low probability of bypassing the quarantine queue \cite{giantsan2024}.

\subsection{Pointer-based Methods}
Pointer-based methods model memory with a focus on pointers by tracking which memory regions are safe for each pointer to access.

SoftBound+CETS \cite{softbound2009,cets2010} ensures spatial and temporal safety through pointer-based bounds checking and identifier metadata associated with pointers. However, SoftBound+CETS introduces high runtime overheads due to expensive metadata propagation and complex logic for checking. EffectiveSan \cite{EffectiveSan2018} enforces type and memory safety using a combination of low-fat pointers, type metadata, and type/bounds check instrumentation. EffectiveSan can detect type and sub-object bounds errors through dynamic type checking, but its temporal safety protection is not as comprehensive as that of CETS. CAMP \cite{CAMP2024} only protects heap memory, validating the memory access of each pointer with boundary checking as well as escape tracking. It also neutralizes dangling pointers with a customized seglist allocator that tracks memory ranges for each allocation. 
Safe Sulong \cite{Sulong2018} compiles C/C++ programs into Java byte code, leveraging the Java virtual machine to take over memory safety detection. Oscar \cite{oscar2017} utilizes page aliases and page-level permissions to achieve heap use-after-free protection. Building upon the philosophy of Oscar, DangZero \cite{dangzero2022} utilizes a privileged backend to mark reserved bits in page tables to implement heap use-after-free detection.
HWASan \cite{hwasan2018} uses ARM’s Top Byte Ignore (TBI) feature to embed an address tag into the top byte of each pointer to identify a memory region. This tag is implicitly propagated to subsequent pointers during pointer assignments. Every load and store instruction raises an exception on a mismatch between the address tag and the memory tag. HWASanIO \cite{HWASanIO2023} adds support for identifying sub-objects based on HWASan, but it incurs significant additional runtime and memory overhead. Similarly, to avoid the extra overhead of tracking pointers, PACMem \cite{PACMem2022} seals the metadata into pointers with ARM's Pointer Authentication (PA) feature.

\section{Conclusions}\label{sec:conclusion}
ASan has become the most popular solution for detecting memory safety violations in C/C++ programs during execution, but it imposes significant runtime overhead. Existing methods for speeding up ASan either fail to adequately eliminate redundant checks or compromise detection capability. To address this issue, this paper presents \textsc{Tech-ASan}, which leverages a novel two-stage check mechanism to effectively reduce ASan's runtime overhead. We also design an efficient optimizer to eliminate redundant checks, which integrates a novel algorithm for removing checks in loops.
Evaluation on the SPEC CPU2006 benchmark demonstrates that \textsc{Tech-ASan} achieves remarkable improvements, reducing runtime overhead by 33.70\% and 17.89\% compared to ASan and ASan\mbox{-}\mbox{-}, respectively. Moreover, under the same redzone setting, \textsc{Tech-ASan} detects 56 fewer false negative cases than ASan and ASan\mbox{-}\mbox{-} when
testing on the Juliet Test Suite.

\begin{acks}
\begin{sloppypar}
The authors would like to appreciate the anonymous reviewers for their feedback, and we thank the constructive suggestions from Yuchen Zhang, Alexander Potapenko, Xinyue Zhou, Cheng Wen, and Jiadong Peng.
This project is supported by the Shenzhen Science and Technology Foundation (General Program, JCYJ20210324093212034), 2022 Guangdong Province Undergraduate University Quality Engineering Project (Shenzhen University Academic Affairs [2022] No. 7), Science and Technology R\&D Program of Shenzhen (20220810135520002),
Guangdong Province Key Laboratory of Popular High Performance Computers 2017B030314073, and Guangdong Province Engineering Center of China-made High Performance Data Computing System.
\end{sloppypar}
\end{acks}

\bibliographystyle{ACM-Reference-Format}
\bibliography{main}
\end{document}